\documentstyle[aps,epsfig]{revtex}
\begin{document}
\draft

\title{Phenomenology of the superconducting state in Sr$_2$RuO$_4$}

\author{M. Sigrist$^{a}$, D. Agterberg$^b$, A. Furusaki$^{a}$,
C. Honerkamp$^b$, K.K. Ng$^{a}$, T.M. Rice$^b$ 
and M.E. Zhitomirsky$^c$}

\address{$^a$ Yukawa Institute for Theoretical Physics, Kyoto University,
Kyoto 606-8502, Japan \\
$^b$ Theoretische Physik, ETH-H\"onggerberg, 8093 Z\"urich, Switzerland \\
$^c$ Department of Physics,   University of Toronto, Toronto, Ontario,
Canada M5S 1A7.}

\maketitle
\begin{abstract}
The symmetry of the superconducting phase of Sr$_2$RuO$_4$ is identified as 
the odd-parity pairing state $ {\bf d} ({\bf k}) = \hat{{\bf z}} (k_x
\pm i k_y) $ based on recent experiments. The experimental evidence for the
so-called orbital dependent superconductivity leads to a single-band
description of superconductivity based on spin fluctuation
mechanism. It is shown that the state $ \hat{{\bf z}} (k_x
\pm i k_y) $ can be stabilized by the spin fluctuation feedback
mechanism analogous to the A-phase in $^3$He and by spin-orbit
coupling effects. 
\end{abstract}

\section{Introduction}

A few years after the discovery of superconductivity in
CuO$_2$-systems, a new oxide-superconductor has been found:
Sr$_2$RuO$_4$ \cite{MAENO}. This system has actually the same layered
perovskite 
structure as La$_2$CuO$_4$, but behaves otherwise very differently.
In its stoichiometric composition Sr$_2$RuO$_4$ is metallic, displays Fermi
liquid behavior and becomes superconducting at the rather low
transition temperature  
$ T_c \approx 1.5$K\cite{MAENO2}. The electronic properties of
Sr$_2$RuO$_4$ are determined by the three $ 4d $-$ t_{2g} $ orbitals 
($ d_{yz},d_{zx},d_{xy} $) of the Ru$^{4+} $-ions which form three
bands that cross the Fermi level \cite{LDA}. As a result we have two
electron-like and one hole-like Fermi surface \cite{DHVA}.
There are clear indications that the superconducting state is
unconventional. The transition temperature is highly sensitive to the
non-magnetic impurities \cite{MACKENZIE} and the NQR-experiments do
not show any 
Hebel-Slichter peak in $ 1/T_1$ \cite{NQR}. Shortly after the discovery of
Sr$_2$RuO$_4$ it has been suggested that this superconductor might
form odd-parity (spin-triplet) Cooper pairs in contrast to the
cuprate-systems \cite{RICE,BASK}. The basis for this claim was
partially the analogy to  
$ ^3 $He (a $p$-wave superfluid) and the presence of ferromagnetism in 
related compounds such as SrRuO$_3$ \cite{RICE}. 

Meanwhile a number of experiments point indeed towards odd-parity
pairing. In $ \mu $SR-experiments the enhancement of the zero-field
relaxation rate in the superconducting state indicates the presence of 
intrinsic magnetism \cite{LUKE}. This occurs in connection with broken
time-reversal symmetry of the superconducting state \cite{SU}. We will 
give in the next section a brief argument on the reason why this
result suggests odd-parity pairing. Even more compelling
evidence is provided by the $^{17}$O NMR Knight
shift data which show that the spin susceptibility is not affected by
the superconducting state for magnetic fields parallel  
to the RuO$_2$-plane. The odd-parity state which is compatible with
all the present data has the form of $ {\bf d} ({\bf k}) = \hat{\bf
z}(k_x \pm i k_y) $. 

In this talk we will discuss the symmetry aspects of the
superconducting state, first ignoring the multi-band structure of this 
compound. In the second part, however, we will analyze the influence
of the different orbitals and give a brief outline of the idea of
orbital dependent superconductivity. Finally, we will examine weak
coupling theories based on ferromagnetic spin fluctuations which could
stabilize the most promising candidate,  $ {\bf d} ({\bf k}) = \hat{\bf
z}(k_x \pm i k_y) $. 

\section{Symmetry of the superconducting state}

The layered perovskite structure of Sr$_2$RuO$_4$ corresponds to the
tetragonal point group $ D_{4h} $. The possible Cooper pairing states
can be classified according to the irreducible representations of 
$D_{4h} $ which include four one-dimensional and one two-dimensional
representation for both even and odd parity \cite{SU}. This determines
the orbital symmetry  
of the even-parity (spin-singlet) states completely. In the case of
odd-parity pairing, however, the Cooper pairs have a spin degree of
freedom  whose
orientation is completely determined, if spin-orbit coupling is
taken into account, i.e. the spin and the crystal lattice orientation are not
independent. In Table I we list both even- and odd-parity states using  
simple basis functions on a cylindrical Fermi surface, i.e. expressed 
in momenta $ {\bf k} $ with $ |{\bf k}| = k_F $. The pair wavefunction 
can be expressed in the standard notation by a scalar function $
\psi({\bf k}) $ for even 
and by a vector function $ {\bf d} ({\bf k}) $ for odd parity pairing
states \cite{SU}.   

It is rather unlikely to find pairing among electrons on different
RuO$_2$-layers as they are well separated. As a consequence, 
even-parity states belonging to the two-dimensional $ E_g
$-representation can be ruled out, because they would require interlayer 
pairing and would actually by symmetry not have any pairing amplitude
within the plane. All other representations possess intra-layer
pairing states. In particular, the
two-dimensional representation $ E_u $ contains the basis states
$ \{ {\bf z} k_x , {\bf z} k_y \}$. In Table II we list the three possible
combinations of these two $ E_u $-basis states. All three states break
a symmetry in addition to the
U(1)-gauge symmetry. The a-phase has broken time reversal symmetry,
while the b- and c-phase lower the crystal field symmetry. Within a
weak-coupling approach the a-phase is the most stable state as we will 
see below, since its gap has no nodes.

Let us now compare these states with some experiments mentioned
above. The only state which breaks time reversal symmetry 
is the a-phase in Table II,  $ {\bf d} ({\bf k}) = \hat{{\bf z}} (k_x
\pm i k_y ) $. Other possible time reversal symmetry breaking states
would involve the complex combinations of two states belonging to different
representations. In general, this would lead to double phase
transitions and only below the second transition the state with broken 
time reversal symmetry would appear. This is in clear contradiction with the
experimental $ \mu $SR-data which show only one phase transition, the onset of 
superconductivity, coincident with the appearance of intrinsic
magnetism, the sign for the violation of time reversal
symmetry \cite{SU}. Thus, this experiment is a strong indication for
the odd-parity 
pairing state $ {\bf d} ({\bf k}) = \hat{{\bf z}}(k_x \pm i k_y) $
which is the two-dimensional analog to the A-phase in superfluid $^3$He.

This interpretation is supported by measurements of the spin susceptibility in 
the superconducting state using $^{17}$O-Knight shift
\cite{ISHIDA}. For odd-parity 
states the uniform susceptibility has the form

\begin{equation}
\chi_{\mu \nu} = \chi_{0} \left[ \delta_{\mu \nu} - \langle (1 -
Y(\hat{{\bf k}}, T)) {\rm Re} d^*_{\mu} (\hat{{\bf k}}) d_{\nu} (\hat{{\bf
k}})  \rangle_{FS} \right]
\label{spinsus}
\end{equation}
where $ Y(\hat{{\bf k}}, T) $ is the angle-dependent Yosida function,
$ \chi_0 $ is the Pauli spin 
susceptibility and $ \langle ... \rangle_{FS} $ denotes the 
average over the Fermi surface \cite{LEGGETT}.
For the state $ \hat{{\bf z}} (k_x \pm i k_y ) $ we obtain,

\begin{equation}
\chi_{\mu \nu} (T) = \chi_{0} \delta_{\mu \nu} 
\left\{ \begin{array}{cl} 
\langle Y(\hat{{\bf k}}, T) \rangle_{FS} & {\bf H} \parallel \hat{{\bf z}} \\
1    & {\bf H} \perp \hat{{\bf z}} \\
\end{array} \right.
\end{equation}
The experiment can only be performed for the field in the basal plane where
it agrees perfectly with the expected result \cite{ISHIDA}.

Based on these experimental findings we can derive the corresponding
Ginzburg-Landau theory which is based on a two-dimensional order
parameter $ \mbox{\boldmath $ \eta $} = ({\eta_x , \eta_y}) $ so that
$ {\bf d} ({\bf 
k}) = \hat{{\bf z}} ({\bf k} \cdot \mbox{\boldmath $ \eta $}) $ . The
free energy has 
the following general form,

\begin{eqnarray}
F\!\!&=&\!\!\!
\int \!d^3 r \biggl[ a(T-T_c) |\mbox{\boldmath $ \eta $}|^2
+ b_1 |\mbox{\boldmath $ \eta $}|^4
+ \frac{b_2}{2} (\eta^{*2}_x \eta^2_y + 
\eta^2_x \eta^{*2}_y) + b_3 |\eta_x|^2 |\eta_y|^2
+ K_1 (|D_x \eta_x |^2 + |D_y \eta_y|^2)
+ K_2 (|D_x \eta_y|^2 + |D_y \eta_x |^2)  \nonumber\\
&&\qquad
+ \{ K_3 (D_x \eta_x)^* (D_y \eta_y)
+ K_4 (D_y \eta_x)^* (D_x \eta_y) +c.c. \}
+ K_5 (|D_z \eta_x|^2 + | D_z \eta_y|^2)
+ \frac{(\nabla \times {\bf A})^2}{8 \pi} \biggr],
\end{eqnarray}
where $ a,b_i $ and $ K_i $ are real coefficients and $ {\bf D} =
\nabla - i 2e {\bf A}/ \hbar c $ is the gauge-invariant gradient
\cite{SU}.
The choice $ b_2 > 0 , b_3 < b_2 $ stabilizes the a-phase with
$\mbox{\boldmath $ \eta 
$} = \eta_0 (1 , \pm i) $ . Based on this free energy Agterberg showed 
that the vortex lattice has square lattice form for fields along the
$z$-axis \cite{DAN}. The orientation of the lattice depends on the
coefficients of the free energy. Recent neutron scattering experiments 
show indeed a square vortex lattice with the main axis orientation
parallel to that of the crystal lattice \cite{FORGAN}. 

\section{Orbital dependent superconductivity}

As we mentioned above the Fermi liquid state is formed by the bands belonging 
to the three $ t_{2g} $-orbitals, $ d_{yz} , d_{zx} $ and $ d_{xy}
$. By symmetry the $ d_{xy} $-band is distinct from the two bands
belonging to the other two orbitals. It can be shown
that the pair scattering between the two types of bands is weak due to 
the special character of the orbitals \cite{AGTER,MAZIN}. Therefore we
may assume that 
the superconductivity is associated with one of the two subsystems while the
other is only participating via the pair component induced by the weak 
pair scattering, which represents a form of proximity effect in the
momentum space. We call this phenomena ``orbital dependent
superconductivity'' (ODS) \cite{AGTER}. 

The superconducting state $ \hat{{\bf z}} (k_x \pm i k_y ) $ opens a
gap over the whole Fermi surface. As a consequence of ODS, however,
this gap is very different in magnitude for the  
Fermi surfaces with intrinsic and induced superconductivity. This
aspect appears, in particular, in thermodynamic quantities such as the
specific heat which 
seems to preserve a large amount of low-energy
density of states (DOS) down to rather low temperatures
\cite{NISHIZAKI}. Theses states close to the  
Fermi level are associated with the orbitals which are not intrinsically
superconducting. Only at rather low
temperature the gap induced on these bands is expected to become
visible in thermodynamic quantities. Experimentally this virtual residual  
DOS has been found to be of the order of 40 - 50 $ \% $ of
the normal state DOS. 

The recent analysis of London penetration depth and coherence length
by Riseman and coworkers led to the further strong evidence for
ODS identifying $ d_{xy} $ as the orbital relevant for
superconductivity \cite{FORGAN}. This yields the value for the
expected residual DOS of about 43 $ \% $  derived from the effective mass
experiments in the de Haas-van Alphen measurements, which is very
consistent with the specific heat data \cite{DHVA}. 
In addition Agterberg has shown that
the $ d_{xy} $-orbital would lead to coefficients of the
Ginzburg-Landau theory in 
the proper range to account for the orientation of the vortex
lattice \cite{DAN}. 

\section{Spin fluctuations and the symmetry of the superconducting
state}

In a recent NMR experiment Imai and coworkers extracted
the uniform spin susceptibility for each of the
orbitals separately from their $^{17} $O-Knight shift data taken in
the normal state over wide temperature range \cite{IMAI}. The
susceptibility associated with the $ d_{xy} $-orbital is considerably
larger than the contributions of the other two bands. Moreover, it is
significantly increasing with lowering temperature, while the other
two bands have a more or less temperature-independent susceptibility. 
This suggests that the tendency towards ferromagnetism is stronger
for the $ d_{xy} $-orbital than the others. In view of the idea of
orbital dependent superconductivity we can simplify the following
discussion by restricting ourselves to
a single band belonging to the relevant orbital ($ d_{xy} $) assuming also
ferromagnetic spin fluctuations as the dominant mechanism for
pairing. Before going into the details of the symmetry analysis of the
resulting  
pairing state, we would like to give here an argument on why
ferromagnetic spin fluctuations are important in
Sr$_2$RuO$_4$. There is a series of ferromagnetic compounds related
to Sr$_2$RuO$_4$, the Ruddlesen-Popper series
Sr$_{n+1}$Ru$_{n}$O$_{3n+1}$, which are multi-layer compounds with $ n
$ as the 
number of RuO$_2$-planes per unit cell. The infinite-layer (3D)
SrRuO$_3$ is a ferromagnet with $T_C \approx 165$K whereby the band
structure calculations give a good understanding within the Stoner
theory \cite{MAZIN2}. For $ n=3 $ one
finds $ T_C \approx 148$K\cite{CAO1} and for $ n=2 $ $ T_C \approx 104$K
(although the ferromagnetism in this latter case is controversial
\cite{CAO2,IKEDA}). This demonstrates the
tendency that with decreasing layer number $ n $ $ T_C$ is
reduced and vanishes finally. Hence, we may consider $ n $ as the
parameter controlling a quantum phase transition between a
ferromagnetic and a magnetically disordered phase. In the schematic
phase diagram depicted in Figure 1 we see that the single-layer compound
Sr$_2$RuO$_4$ would lie very near to the quantum critical point so that
ferromagnetic spin fluctuations play an important role. We may argue that
the moderately enhanced Wilson ratio $ R_W \sim 1.4 $ would contradict this
conclusion. Note, however, that the two-dimensionality plays an
important role in reducing $ R_W $ as pointed out recently by Julian and
coworkers \cite{LONZARICH}. In two dimensions it should diverge as $
R_W \propto 
(1-a)^{-1/2} $ as the quantum phase transition is approached
(controlling parameter $ a \to
1 $) which is weaker than in  three dimensions where
$ R_w \propto - {\rm ln}(1-a)/(1-a) $. Therefore the comparison with
other (three-dimensional) nearly ferromagnetic systems could be
misleading. Moreover there are other mechanisms renormalizing the
mass. 

\subsection{Phenomenological model}

Now we discuss a single band model for Cooper pairing assuming that
the (relevant) $ d_{xy} $-band has 
cylindrical symmetry. The Hamiltonian has the form,
\begin{equation}
{\cal H} =
 \sum_{k,s} \epsilon_k c^{\dag}_{{\bf k} s} c_{{\bf k} s}
+ \frac{1}{2} \sum_{{\bf k}, {\bf k}'} \sum_{s_1, ... ,s_4}
 V_{{\bf k},{\bf k}'; s_1 s_2 s_3 s_4}
c^{\dag}_{{\bf k} s_1} c^{\dag}_{- {\bf k} s_2}
 c_{-{\bf k}' s_3} c_{{\bf k}' s_4},
\end{equation}
where $ \epsilon_k $ denotes the electron band energy measured from
the Fermi energy and $ c^{\dag}_{{\bf k}s} $ and $ c_{{\bf k}s} $ are
the Fermion creation and annihilation operators. The effective pairing
interaction is mediated by the spin fluctuations (paramagnon exchange)
which we describe here by the static susceptibility $ \chi_{\mu \nu}
({\bf q}) $ , 

\begin{equation}
V_{{\bf k},{\bf k}'; s_1 s_2 s_3 s_4} = -\frac{I^2}{4} \sum_{\mu, \nu}
\chi_{\mu \nu} ({\bf k} - {\bf k}') \sigma^{\mu}_{s_1 s_4}
\sigma^{\nu}_{s_2 s_3}
\end{equation}
where $ I $ is an interaction constant. We ignore the dynamical part of
the spin fluctuations and adopt a weak coupling approach where the
interaction is finite in a certain range around the Fermi energy. The
corresponding cutoff frequency  $ \omega_c $ limiting the attractive region 
is not so easy to define, but is sometimes brought into connection
with the largest paramagnon frequency,  $ \gamma q_c (1-a)/\chi_{0}$
with a cutoff wave vector  around $ 2 k_F $ and $ 
\gamma $ a phenomenological parameter \cite{FAY}. For the 
static susceptibility we use for simplicity the small-$q$
approximation $ \chi_{\mu \nu} ({\bf 
q}) = \chi_0 \delta_{\mu \nu} / (1- a + c q^2) $ where $c $ is the 
spin stiffness ($ q^2 = q^2_x + q^2_y $) \cite{MORIYA}. 
The factor $ (1-a)^{-1} $ describes the Stoner enhancement. This
approximation is certainly an oversimplification of the real situation
and a more realistic form based on band structure data may be found in 
Ref.\cite{MAZIN}. This simplification would, however, not invalidate
our further discussion, since we will concentrate on aspects related
to symmetry of the states by comparing the classified states in Table I 
and II. The momentum structure should affect all states essentially in
the same way. 

The standard BCS paring meanfield scheme leads to the self-consistence 
equation given by
\begin{equation}
d^{\alpha}({\bf k}) = - \frac{I^2}{4N} \sum_{{\bf k'}, \beta}
\left[\sum_{\mu} \chi_{\mu \mu}({\bf k} - {\bf k}')
 \delta_{\alpha \beta} - \chi_{\alpha \beta}({\bf k} - {\bf k}') 
- \chi_{\beta \alpha}({\bf k} - {\bf k}') \right]
\frac{d^{\beta}({\bf k}')}{2 E_{{\bf k}'}}
\tanh\left(\frac{E_{{\bf k}'}}{2 k_B T}\right)
\end{equation}
with $ E_{{\bf k}} = [ \epsilon_k + |{\bf d}({\bf k})|^2]^{1/2} $ and 
excluding non-unitary pairing states, i.e. we impose the condition $ {\bf d}^*
\times {\bf d} = 0 $.

\subsection{Spin rotation symmetric case}

In the absence of spin-orbit coupling $ \chi_{\mu 
\nu} ( {\bf q}) = \chi(q) \delta_{\mu \nu} $ as given above. Assuming
a small cutoff energy we can expand $ \chi({\bf k} - {\bf k}') $ in Legendre 
polynomials $ P_l (\theta) $  where $ \theta $ denotes the angle
between $ {\bf k} $ and $ {\bf k} ' $ for momenta lying on the Fermi
surface. The lowest and only relevant component is $ P_{l=1} (\theta)
= \cos (\theta) / \sqrt{\pi} $ which yields the self-consistence equation 

\begin{equation}
d^{\alpha} ({\bf k}) = \frac{g}{N} \sum_{{\bf k}'} \frac{ {\bf k} \cdot 
{\bf k}'}{k^2_F} \frac{d^{\alpha} ({\bf k}') }{2 E_{{\bf k}'}} 
{\rm tanh}\left(\frac{E_{{\bf k}'}}{2 k_B T}
\right) 
\label{self}
\end{equation}
with the coupling constant $ g \approx (I^2 \chi_0 /8 k_F) / \sqrt{c
 (1-a)} $. The superconducting transition temperature is obtained
from the linearized self-consistence equation,

\begin{equation}
k_B T_c = 1.14 \omega_c {\rm exp}( - 1/ N(0) g)
\label{tc}
\end{equation}
where $ N(0) $ is the DOS of the $ d_{xy} $-band at the Fermi level. 
In this simplified calculation we have, however, ignored the effect of 
the spin fluctuation on the renormalization of the quasiparticles
forming the Cooper pairs. This leads to an additional renormalizing
factor in the exponent of Eq.(\ref{tc}), $ {\rm exp}(-(1+ \lambda)/N(0) g) $ 
where $ \lambda $ is approximately proportional to $ g $ for the
contribution due to the spin fluctuations \cite{FAY,MAZIN}. Note, however,
that the multi-band structure makes this analysis 
somewhat more complicated than in the pure single-band case \cite{MAZIN}. 

The transition temperature $ T_c $ given in Eq.(\ref{tc}) is the same
for any state of the form $ {\bf d} ({\bf k}) = \sum_{\alpha, \mu}
d_{\alpha \mu} \hat{{\bf n}}_{\alpha} k_{\mu} $  where $ \hat{{\bf
n}}_{\alpha} $ is the unit vector in $ \alpha $-direction in the basis 
for the $ {\bf d} $-vectors. In order to find the states which are
actually stable in the self-consistence equation we have to go beyond
the linearized form. Already the Ginzburg-Landau theory which is
easily derived from Eq.(\ref{self}) using $ d_{\alpha \mu} $ as order
parameters, is sufficient.
\begin{equation}
F = f_0 \left\{ \sum_{\alpha, \mu} \ln\left(\frac{T}{T_c}\right)
|d_{\alpha\mu}|^2 + \frac{b}{16} \sum_{\alpha,\beta,\mu,\nu}
\left[ |d_{\alpha \mu}|^2 | d_{\beta \nu}|^2
 + d^*_{\alpha \mu} d_{\alpha \nu} d^*_{\beta\mu} d_{\beta \nu}
 + d^*_{\alpha \mu} d_{\alpha \nu} d^*_{\beta\nu} d_{\beta \mu}
 \right]
\right\}
\label{glf}
\end{equation}
with $ b = 7 \zeta(3)/8 (\pi k_B T_c)^2 $. By examination we find that 
states which have gaps $ |{\bf d} ({\bf k}) |^2 $ without nodes are
more stable than those with nodes. All states belonging to the
one-dimensional representations, $ A_{1u}, A_{2u} , B_{1u} , B_{2u} $ 
and the a-phase of the $ E_u $-representation, as listed
in Table I and II, have nodeless gaps. Within this free energy analysis 
they are degenerate. Thus the spin symmetric weak-coupling theory
results in six pairing states of the same condensation energy, a
property specific to two 
dimension. In three dimensions there is only one (non-degenerate)
state of this kind, the BW-state, which forms the B-phase of
superfluid $^3$He \cite{LEGGETT}. 
 
We will now investigate
further mechanisms which would favor one state over the others.
One such mechanism is based on the feedback effects of the
superconducting state on the spin fluctuations which is well known
to stabilize the A-phase in $^3$He. Let us give here a simplified
approach to this mechanism which, however, contains all the essential
physics, following Leggett \cite{LEGGETT}. 
The presence of
superconductivity modifies the spin susceptibility entering the
pairing interaction. 

\begin{equation} \begin{array}{ll}
\chi_{\mu \nu} (q) & = \chi_N (q) \delta_{\mu \nu} + \delta \chi_{\mu
\nu} (q) \\ & \\
& = \chi_N (q) \delta_{\mu \nu} + f(q) \delta \chi_{0 \mu
\nu}.  \\
\end{array}
\end{equation}
We introduce the form factor $ f(q) $ for the correction of
the (uniform) susceptibility \cite{LEGGETT,VOLLHARDT} which is given by
Eq.(\ref{spinsus}), 

\begin{equation}  
\delta \chi_{0 \mu \nu} = \chi_0 2 b \langle {\rm Re} d^*_{\mu} ({\bf
k}) d_{\nu}({\bf k}) \rangle_{FS} + O(|{\bf d}|^4)
\end{equation}
where we have expanded $ \delta \chi_{0 \mu \nu} $ in terms of $ {\bf d} $
close to $ T= T_c $. Inserting this expression as a correction to the
interaction into
the self-consistence equation we can easily derive an additional
fourth-order term to the Ginzburg-Landau free energy in Eq.(\ref{glf}) 
of the form,
\begin{equation}
\delta F = \kappa \sum_{\alpha, \beta, \mu, \nu}
({\rm Re}\,d^*_{\alpha \mu} d_{\beta \mu} )
\{\delta_{\alpha \beta} 
\sum_{\alpha'} d^*_{\alpha' \nu} d_{\alpha' \nu}
 - 2 ({\rm Re}\, d^*_{\alpha\nu} d_{\beta \nu} ) \} ,
\end{equation}
where $ \kappa $ is a positive constant
\cite{LEGGETT,VOLLHARDT}. Inserting the different 
nodeless states we find that $ \hat{{\bf z}} ( k_x \pm i k_y ) $ is
energetically favored over all the other states, in close
analogy to the A-phase of $^3$He. 

\subsection{The effect of spin-orbit coupling}

The spin fluctuation feedback mechanism stabilized the a-phase in the
fourth order term of the free energy expansion. In second order there
is still complete degeneracy among all the states due to the spin 
rotation symmetry. We now discuss the effect of spin-orbit coupling
which lifts this degeneracy. With spin-orbit coupling the spin
susceptibility entering into the  
pairing interaction has reduced symmetry in the sense that it needs to 
be a scalar only under simultaneous rotations of both spin and orbital 
part and not anymore under their separate transformation. On the
lowest level of  
expansion equivalent to the term (proportional to the first Legendre
polynomial) given in Eq.(\ref{self}) we find
the general form invariant under cylindrical symmetry,

\begin{equation}
\chi_{\mu \nu} ({\bf k} - {\bf k}') = \chi^{(1)}_{\mu} \delta_{\mu
\nu} {\bf k} \cdot {\bf k}' + \chi^{(2)}_{\mu \nu} k_{\mu} k'_{\nu}
\end{equation}
which we approximate by essentially three phenomenological parameters,
$ \chi^{(1)}_{x,y} = g_1 $, $ \chi^{(1)}_z = g_2 $ and $ \chi^{(2)}_{xy} 
= - \chi^{(2)}_{yx} = g_3 $. In the linearized gap equation this 
susceptibility leads to
\begin{equation}
v d^{\alpha} ({\bf k}) =
\int \frac{d \Omega'}{\pi k^2_F}
\sum_{\beta}\left[
(2 g_1 + g_2 - 2 \chi^{(1)}_{\alpha})
 {\bf k} \cdot {\bf k}'\delta_{\alpha \beta}
 - g_3 (k_{\alpha} k'_{\beta} - k_{\beta} k'_{\alpha} ) \right]
d^{\beta} ({\bf k}')
\label{gap}
\end{equation}
with $ k_B T_c = 1.14 \omega_c {\rm exp} ( - 4/ I^2 N(0) v) $ and $ v
$ is an eigenvalue in this equation. In Table III we give a list of the
eigenvalues and eigenstates. The states separate into three subsets of 
doubly degenerate eigenvalues (transition temperature) according to their total
angular momentum $ J_z $. The stable state is decided by the choice of
the parameters $ g_{1-3} $ ($ |g_3| \sim | g_1 - g_2 | $). Therefore, if
the spin fluctuations are enhanced for spin 
orientations in the basal plane the 
state $ \hat{{\bf z}} (k_x \pm i k_y ) $ can reach the highest transition
temperature. Indeed recent experiments by Mukuda et al. show that the
ferromagnetic spin fluctuations are apparently stronger in-plane than
out-of-plane \cite{MUKUDA}. Whether this is the case for the $d_{xy}
$-orbital remains to be investigated.

\section{Conclusions}

The interpretation of various experiments led to the conclusion that
the most likely superconducting state 
of Sr$_2$RuO$_4$ is given by the time reversal symmetry breaking
pairing state of  the form $ {\bf d} ({\bf k}) = \hat{{\bf z}} (k_x
\pm i k_y) $ reminding of the A-phase of superfluid $^3$He. This state is
compatible with all currently available experimental data. Also recent 
investigations on
Josephson contacts between Sr$_2$RuO$_4$ and Pb should be included here
\cite{JIN}, although their interpretation is not
completed yet \cite{TANAKA,HONERKAMP}. 
Furthermore, there is good reason for the assumption that
superconductivity can be mainly associated with the $ d_{xy} $-band,
while the other two orbitals participate only passively via
proximity-induced pairing. This could, in particular, account for the
large apparent residual DOS in the superconducting state. It seems
rather unlikely that a 
non-unitary state as proposed earlier to explain the large residual
DOS \cite{ZHITO} is realized here, since such a state is very 
difficult to stabilize unlike the unitary state $ \hat{{\bf z}} (k_x
\pm i k_y) $ \cite{MAZIN}. We have shown that a spin 
fluctuation based mechanism favors the state $ \hat{{\bf z}} (k_x
\pm i k_y) $. There are basically two stabilizing mechanisms: (1) the
spin fluctuation feedback analogous to $^3$He and (2) the spin-orbit
coupling effects enhancing the ferromagnetic spin fluctuation with spin
orientations parallel to RuO$_2$-plane. 

In conclusion, we would like to emphasize that Sr$_2$RuO$_4$ is probably the
first odd-parity superconductor emerging out of a strongly correlated
electron system which is a clear Fermi liquid. In the case of the
heavy Fermion superconductors UPt$_3$  
and UBe$_{13}$, two other candidates for odd-parity pairing, the
situation is considerably more complicated. 
Most experimental data for Sr$_2$RuO$_4$ show that the properties of the
superconductor can be investigated in a controlled and clean
way. Therefore, Sr$_2$RuO$_4$ may in future become the textbook example for the
study and presentation of unconventional superconductivity. The fact 
that the superconducting order parameter has two components, yields
also a large space for unusual phenomena \cite{SU,TANIGUCHI}.
These include the formation of domains separated by domain walls,
non-axial vortices or various collective modes, to mention only a few.

\vskip 0.4 cm

We are grateful for stimulating discussions with G. Lonzarich,
H. Kontani, I. Mazin, 
S. Murakami, N. Nagaosa, Y. Okuno, and K. Ueda, and with various experimental
groups, in particular, Y. Maeno and his collaborators. This work has
been financially supported by the Ministry of Education, Science and
Culture of Japan.

\narrowtext

\begin{table}
\begin{center}

\begin{tabular}
{|@{\hspace{\tabcolsep}\extracolsep{\fill}}c|c||c|c|}
\hline
$ \Gamma $ & $ \psi({\bf k}) $ & $ \Gamma $ & $ {\bf d}({\bf k}) $ \\
\hline
$ A_{1g} $ & 1 & $ A_{1u} $ & $ \hat{{\bf x}} k_x + \hat{{\bf y}} k_y
$ \\
$ A_{2g} $ & $ k_x k_y (k^2_x -k^2_y) $ & $ A_{2u} $ & $ \hat{{\bf x}} k_y
- \hat{{\bf y}} k_x $ \\
$ B_{1g} $ & $ k^2_x -k^2_y $ & $ B_{1u} $ & $ \hat{{\bf x}} k_x
- \hat{{\bf y}} k_y $ \\
$ B_{2g} $ & $ k_x k_y $ & $ B_{2u} $ & $ \hat{{\bf x}} k_y
+ \hat{{\bf y}} k_x $ \\
$ E_g $ & - & $ E_u $ & $ \{ \hat{{\bf z}} k_x , \hat{{\bf z}} k_y \}
$ \\
\hline
\end{tabular}
\caption{List of possible pairing states for the tetragonal point
group $ D_{4h} $ with even and odd parity.}
\end{center}
\end{table}

\begin{table}
\begin{center}
\begin{tabular}
{|@{\hspace{\tabcolsep}\extracolsep{\fill}}c|c|}
\hline
a-phase & $ {\bf d} ({\bf k}) = \hat{{\bf z}} ( k_x \pm i k_y) $ \\
b-phase & $ {\bf d} ({\bf k}) = \hat{{\bf z}} ( k_x \pm k_y) $ \\
c-phase & $ {\bf d} ({\bf k}) = \hat{{\bf z}} k_x, \hat{{\bf z}}k_y  $
\\ \hline
\end{tabular}
\caption{The three phases of the $ E_u $-representation with basis $
\{ \hat{{\bf z}} k_x , \hat{{\bf z}} k_y \} $.}
\end{center}
\end{table}

\begin{table}
\begin{center}
\begin{tabular}
{|@{\hspace{\tabcolsep}\extracolsep{\fill}}ccc|}
\hline
$ {\bf d} ({\bf k}) $  & $ J_z $ & $ v $ \\
\hline
$ \hat{{\bf x}} k_x + \hat{{\bf y}} k_y $ & 0 & $ 2 (g_2 - g_3) $ \\
$ \hat{{\bf x}} k_y - \hat{{\bf y}} k_x $ & 0 &  $ 2 (g_2 - g_3) $ \\
$ \hat{{\bf x}} k_x - \hat{{\bf y}} k_y $ & $ \pm 2 $ &  $ 2 (g_2 + g_3) $ \\
$ \hat{{\bf x}} k_y + \hat{{\bf y}} k_x $ & $ \pm 2 $ & $ 2 (g_2 + g_3) $ \\
$ \{ \hat{{\bf z}} k_x , \hat{{\bf z}} k_y \} $ & $ \pm 1 $ & $ 2 g_1 $ 
\\ \hline
\end{tabular}  
\vskip 0.1 cm
\caption{The eigenvalues and eigenstates of the gap equation
Eq.(\ref{gap}) with their total angular momentum $ J_z $. }
\end{center}
\end{table} 

\widetext

\vspace*{3mm}
\begin{figure}
\begin{center}
\epsfxsize=105mm
\epsffile{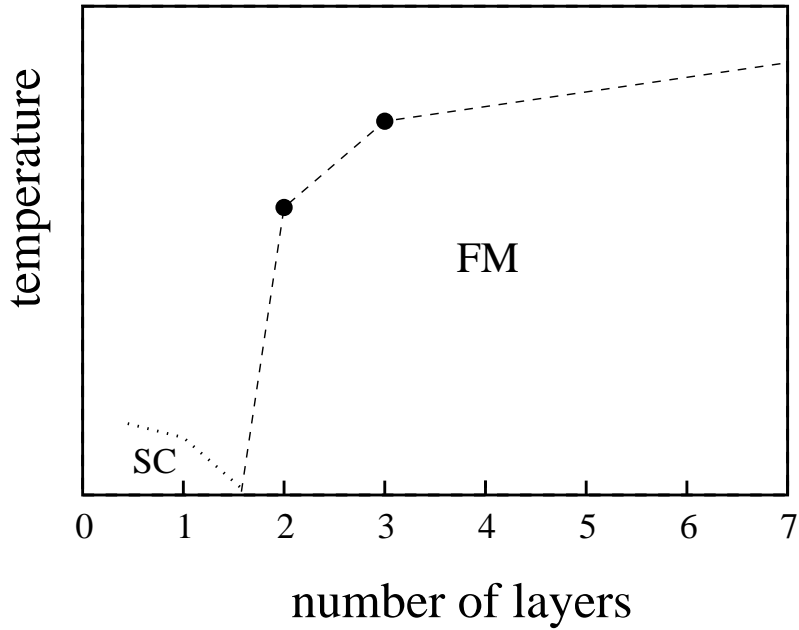}
\vspace*{3mm}
\caption[]{Schematic phase diagram of the ferromagnetic and
superconducting members of the Ruddlesen-Popper series. The number of
layers is the parameter which determines the quantum phase transition
between the two phases. (FM: ferromagnetic,; SC: superconducting).}
\end{center}
\end{figure}


\begin{thebibliography}{99}

\bibitem{MAENO} Y. Maeno, H. Hashimoto, K. Yoshida, S. Nishizaki, T.
Fujita, J.G. Bednorz and F. Lichtenberg, Nature {\bf 372}, 532 (1994).

\bibitem{MAENO2}  Y. Maeno, Physica C {\bf 282 - 287}, 206 (1997).

\bibitem{LDA} T. Oguchi, Phys. Rev. B{\bf 51}, 1385 (1995);
D.J. Singh, Phys. Rev. B{\bf 52}, 1358 (1995).

\bibitem{DHVA} A.P. Mackenzie, S.R. Julian, A.J. Diver, G.G.
Lonzarich, Y. Maeno, S. Nishizaki and T. Fujita, Phys. Rev. Lett. {\bf
76}, 3786 (1996).

\bibitem{MACKENZIE} A.P. Mackenzie, R.K.W. Haselwimmer, A.W. Tyler,
G.G. Lonzarich, Y. Mori, S. Nishizaki and Y. Maeno,
Phys. Rev. Lett. {\bf 80}, 161 (1998).

\bibitem{NQR} K. Ishida, Y. Kitaoka, K. Asayama, S. Ikdea,
S. Nishizaki, Y. Maeno, K. Yoshida and T. Fujita, Phys. Rev. B {\bf
56}, R505 (1997).

\bibitem{RICE} T.M. Rice and M. Sigrist: J. Phys., Condens. Matter
{\bf 7}, L643 (1995).

\bibitem{BASK} G. Baskaran, Physica B {\bf 223 \& 224},
490 (1996). 

\bibitem{LUKE} G.M. Luke, Y. Fudamoto, K.M. Kojima, M.I. Larkin,
J. Merrin, B. Nachumi, Y.J. Uemura, Y. Maeno, Z.Q. Mao, Y. Mori,
H. Nakamura and M. Sigrist, Nature {\bf 394}, 558 (1998).

\bibitem{SU} M. Sigrist and K. Ueda, Rev. Phys. Mod. {\bf 63}, 239 (1991).

\bibitem{ISHIDA} K. Ishida, H. Mukuda, Y. Kitaoka, K. Asayama,
Z.Q. Mao, Y. Mori and Y. Maeno, Nature {\bf 396}, 658 (1998)..

\bibitem{MAZIN2} I.I. Mazin and D.J. Singh, Phys. Rev. {\bf B56}, 2556 
(1997).

\bibitem{LEGGETT} A.J. Leggett, Rev. Mod. Phys. {\bf 47}, 331 (1975).


\bibitem{AGTER} D.F. Agterberg, T.M. Rice and M. Sigrist,
Phys. Rev. Lett. {\bf 78}, 3374 (1997).


\bibitem{MAZIN} I.I. Mazin and D.J. Singh, Phys. Rev. Lett. {\bf 79},
733 (1997); Proceedings of SNS'97 Conference, Cape Cod (1997).

\bibitem{NISHIZAKI} S. Nishizaki, Y. Maeno, S. FArner, S. Ikeda and
T. Fujita, J. Phys. Soc. Jpn. {\bf 67}, 560 (1998).

\bibitem{DAN} D.F. Agterberg, Phys. Rev. Lett. {\bf 80}, 5184 (1998).

\bibitem{FORGAN} T. M. Riseman, P. G. Kealey, E. M.
Forgan, A. P. Mackenzie, L. M. Galvin, A. W. Tyler, S. L. Lee, C. Ager,
D. McK Paul, C. M. Aegerter, R. Cubitt, Z. Q. Mao, S. Akima, and Y.
Maeno, Nature {\bf 396}, 242 (1998).

\bibitem{IMAI} T. Imai, A.W. Hunt, K.R. Thurber and F.C. Chou,
Phys. Rev. Lett. {\bf 81}, 3006 (1998).

\bibitem{CAO1} G. Cao, S.K. McCall, J.E. Crow and R.P. Guertin,
Phys. Rev. {\bf B56}, R5740 (1997).

\bibitem{CAO2} G. Cao, S. McCall and J.E. Crow, Phys. Rev. {\bf B55},
R672 (1998).

\bibitem{IKEDA} S. Ikeda, Y. Maeno and T. Fujita, Phys. Rev. {\bf 57}, 
978 (1998).

\bibitem{LONZARICH} S.R. Julian, A.P. Mackenzie, G.G. Lonzarich,
C. Bergemann, R.K.W. Haselwimmer, Y. Maeno, S. Nishizaki, A.W. Tyler,
S. Ikeda and T. Fujita, preprint.

\bibitem{FAY} D. Fay and J. Appel, Phys. Rev. {\bf B22}, 3173 (1980).

\bibitem{MORIYA} T. Moriya, {\it Spin Fluctuations in Itinerant
Electron Magnetism}, Springer (1985).

\bibitem{VOLLHARDT} D. Vollhardt and P. W\"olfle, {\it The Superfluid
Phases of Helium 3}, Taylor \& Francis (1990). 

\bibitem{MUKUDA} H. Mukuda, K. Ishida, Y. Kitaoka, K. Asayama, Z. Mao, 
Y. Mori and Y. Maeno, preprint.

\bibitem{JIN} R. Jin, Y. Zadorozhny, Y. Liu, Y. Mori, Y. Maeno,
D.G. Schlom and F. Lichtenberg, preprint. 

\bibitem{TANAKA} M. Yamashiro, Y. Tanaka and S. Kashiwaya,
J. Phys. Soc. Jpn. {\bf 67}, 3364 (1998).

\bibitem{HONERKAMP} C. Honerkamp and M. Sigrist, 
Prog. Theor. Phys. {\bf 100}, 53 (1998).

\bibitem{ZHITO} K. Machida, M. Ozaki and T. Ohmi,
J. Phys. Soc. Jpn. {\bf 65} (1996), 3720;
M. Sigrist and M.E. Zhitomirsky,
J. Phys. Soc. Jpn. {\bf 65} (1996), 3452.

\bibitem{TANIGUCHI} M. Sigrist, C. Honerkamp, D. Agterberg, T.M. Rice, 
M.E. Zhitomirsky and A. Furusaki, Proceedings of the Taniguchi
Symposium on Strongly correlated electron systems, preprint. 

\end{thebibliography}
\end{document}